# Fairness and Bias in Artificial Intelligence: A Brief Survey of Sources, Impacts, and Mitigation Strategies

Emilio Ferrara

Thomas Lord Department of Computer Science, USC Viterbi School of Engineering

University of Southern California

## Abstract

The significant advancements in applying Artificial Intelligence (AI) to healthcare decision-making, medical diagnosis, and other domains have simultaneously raised concerns about the fairness and bias of AI systems. This is particularly critical in areas like healthcare, employment, criminal justice, credit scoring, and increasingly, in generative AI models (GenAI) that produce synthetic media. Such systems can lead to unfair outcomes and perpetuate existing inequalities, including generative biases that affect the representation of individuals in synthetic data. This survey paper offers a succinct, comprehensive overview of fairness and bias in AI, addressing their sources, impacts, and mitigation strategies.

We review sources of bias, such as data, algorithm, and human decision biases—highlighting the emergent issue of generative AI bias where models may reproduce and amplify societal stereotypes. We assess the societal impact of biased AI systems, focusing on the perpetuation of inequalities and the reinforcement of harmful stereotypes, especially as generative AI becomes more prevalent in creating content that influences public perception. We explore various proposed mitigation strategies, discussing the ethical considerations of their implementation and emphasizing the need for interdisciplinary collaboration to ensure effectiveness.

Through a systematic literature review spanning multiple academic disciplines, we present definitions of AI bias and its different types, including a detailed look at generative AI bias. We discuss the negative impacts of AI bias on individuals and society and provide an overview of current approaches to mitigate AI bias, including data pre-processing, model selection, and post-processing. We emphasize the unique challenges presented by generative AI models and the importance of strategies specifically tailored to address these.

Addressing bias in AI requires a holistic approach, involving diverse and representative datasets, enhanced transparency and accountability in AI systems, and the exploration of alternative AI paradigms that prioritize fairness and ethical considerations. This survey contributes to the ongoing discussion on developing fair and unbiased AI systems by providing an overview of the sources, impacts, and mitigation strategies related to AI bias, with a particular focus on the emerging field of generative AI.

# I. INTRODUCTION

The growing use of AI systems has intensified discussions around fairness and bias in artificial intelligence, as potential biases and discrimination become more apparent. This survey examines the sources, impacts, and mitigation strategies related to fairness and bias in AI. Multiple studies have identified biases against certain groups in AI systems, such as the facial recognition systems studied by Buolamwini and Gebru (2018), and hiring algorithms examined by Dastin (2018) and Kohli (2020). These biases can perpetuate systemic discrimination and inequality, with detrimental effects on individuals and communities in areas like hiring, lending, and criminal justice (O'Neil, 2016; Eubanks, 2018; Barocas and Selbst, 2016; Kleinberg et al., 2018).

Researchers and practitioners have proposed various mitigation strategies, such as improving data quality (Gebru et al., 2021) and designing explicitly fair algorithms (Berk et al., 2018; Friedler et al., 2019; Yan et al., 2020). This paper provides a comprehensive overview of the sources and impacts of bias in AI, examining data, algorithmic, and user biases, along with their ethical implications. It surveys current research on mitigation strategies, discussing their challenges, limitations, and the significance of interdisciplinary collaboration.

The importance of fairness and bias in AI is widely recognized by researchers, policymakers, and the academic community (Kleinberg et al., 2017; Caliskan et al., 2017; Buolamwini and Gebru, 2018; European Commission, 2019; Schwartz et al., 2022; Ferrara, 2023). This survey paper delves into the complex and multifaceted issues surrounding fairness and bias in AI, covering the sources of bias, their impacts, and proposed mitigation strategies. Overall, the paper aims to contribute to ongoing efforts to develop more responsible and ethical AI systems by shedding light on the sources, impacts, and mitigation strategies of fairness and bias in AI.

# II. SOURCES OF BIAS IN AI

Artificial intelligence (AI) has the potential to revolutionize many industries and improve people's lives in countless ways. However, one of the major challenges facing the development and deployment of AI systems is the presence of bias. Bias refers to the systematic errors that occur in decision-making processes, leading to unfair outcomes. In the context of AI, bias can arise from various sources, including data collection, algorithm design, and human interpretation. Machine learning models, which are a type of AI system, can learn and replicate patterns of bias present in the data used to train them, resulting in unfair or discriminatory outcomes. In this section, we will explore the different sources of bias in AI, including data bias, algorithmic bias, and user bias, and examine real-world examples of their impact.

### DEFINITION OF BIAS IN AI AND ITS DIFFERENT TYPES

Bias is defined as a systematic error in decision-making processes that results in unfair outcomes. In the context of AI, bias can arise from various sources, including data collection, algorithm design, and human interpretation. Machine learning models, which are a type of AI system, can learn and replicate patterns of bias present in the data used to train them, resulting in unfair or discriminatory outcomes. It is important to identify and address bias in AI to ensure that these systems are fair and equitable for all users. In the next sections, we will explore the sources, impacts, and mitigation strategies of bias in AI in more detail.

## SOURCES OF BIAS IN AI, INCLUDING DATA BIAS, ALGORITHMIC BIAS, AND USER BIAS

Sources of bias in AI can arise from different stages of the machine learning pipeline, including data collection, algorithm design, and user interactions. This survey discusses the different sources of bias in AI and provides examples of each type, including data bias, algorithmic bias, and user bias (Selbst et al., 2016; Crawford & Calo, 2016).

Data bias occurs when the data used to train machine learning models is unrepresentative or incomplete, leading to biased outputs. This can happen when the data is collected from biased sources, or when the data is incomplete, missing important information, or contains errors. Algorithmic bias, on the other hand, occurs when the algorithms used in machine learning models have inherent biases that are reflected in their outputs. This can happen when algorithms are based on biased assumptions or when they use biased criteria to make decisions. User bias occurs when the people using AI systems introduce their own biases or prejudices into the system, consciously or unconsciously. This can happen when users provide biased training data or when they interact with the system in ways that reflect their own biases.

To mitigate these sources of bias, various approaches have been proposed, including dataset augmentation, bias-aware algorithms, and user feedback mechanisms. Dataset augmentation involves adding more diverse data to training datasets to increase the representativeness and reduce bias. Bias-aware algorithms involve designing algorithms that consider different types of bias and aim to minimize their impact on the system's outputs. User feedback mechanisms involve soliciting feedback from users to help identify and correct biases in the system.

Research in this area is ongoing, with new approaches and techniques being developed to address bias in AI systems. It is important to continue to investigate and develop these approaches to create AI systems that are more equitable and fairer for all users.

## REAL-WORLD EXAMPLES OF BIAS IN AI

There have been numerous examples of bias in AI systems across various industries, from healthcare to criminal justice. One well-known example is the COMPAS system used in the United States criminal justice system, which predicts the likelihood of a defendant reoffending. A study by ProPublica found that the system was biased against African-American defendants, as they were more likely to be labeled as high-risk even if they had no prior convictions. Another study found similar biases in a similar system used in the state of Wisconsin (Angwin et al., 2016).

In healthcare, an AI system used to predict patient mortality rates was found to be biased against African-American patients. A study conducted by Obermeyer et al. (2019) found that the system was more likely to assign higher-risk scores to African-American patients, even when other factors, such as age and health status, were the same. This bias can lead to African-American patients being denied access to healthcare or receiving subpar treatment.

Another example of bias in AI systems is the facial recognition technology used by law enforcement agencies. A study by the National Institute of Standards and Technology (NIST) found that facial recognition technology was significantly less accurate for people with darker skin tones, leading to higher rates of false positives (Schwartz et al., 2022). This bias can have serious consequences, such as wrongful arrests or convictions.

Finally, with the rise of generative AI systems (GenAI), the risk of harmful biases increases (Ferrara, 2023; Ferrara, 2023b). A striking instance of GenAI bias was reported, where text-to-image models like StableDiffusion, OpenAI's DALL-E, and Midjourney, exhibited racial and stereotypical biases in their outputs (Nicoletti & Bass, 2023).

When prompted to generate images of CEOs, these models predominantly produced images of men, reflecting gender bias. This bias mirrors the underrepresentation of women in CEO positions in the real world. Furthermore, when prompted to generate images of criminals or terrorists, the models' output overwhelmingly more people of color.

This incident underscores the risk of generative AI perpetuating societal biases. GenAI models trained on internet-sourced images likely suffer from this bias, as the data mirrored existing disparities. This example highlights the critical need for diverse and balanced training datasets in AI development to ensure fair and representative outputs from generative models.

These examples illustrate the serious consequences of bias in AI systems and the need for careful evaluation and mitigation strategies to address such biases.

| Type of Bias | Description | Examples |
|---|---|---|
| **Sampling Bias** | Occurs when the training data is not representative of the population it serves, leading to poor performance and biased predictions for certain groups. | A facial recognition algorithm trained mostly on white individuals that performs poorly on people of other races. |
| **Algorithmic Bias** | Results from the design and implementation of the algorithm, which may prioritize certain attributes and lead to unfair outcomes. | An algorithm that prioritizes age or gender, leading to unfair outcomes in hiring decisions. |
| **Representation Bias** | Happens when a dataset does not accurately represent the population it is meant to model, leading to inaccurate predictions. | A medical dataset that under-represents women, leading to less accurate diagnosis for female patients. |
| **Confirmation Bias** | Materializes when an AI system is used to confirm pre-existing biases or beliefs held by its creators or users. | An AI system that predicts job candidates' success based on biases held by the hiring manager. |
| **Measurement Bias** | Emerges when data collection or measurement systematically over- or under-represents certain groups. | A survey collecting more responses from urban residents, leading to an under-representation of rural opinions. |
| **Interaction Bias** | Occurs when an AI system interacts with humans in a biased manner, resulting in unfair treatment. | A chatbot that responds differently to men and women, resulting in biased communication. |
| **Generative Bias** | Occurs in generative AI models, like those used for creating synthetic data, images, or text. Generative bias emerges when the model's outputs disproportionately reflect specific attributes, perspectives, or patterns present in the training data, leading to skewed or unbalanced representations in generated content. | A text generation model trained predominantly on literature from Western authors may over-represent Western cultural norms and idioms, under-representing or misrepresenting other cultures. Similarly, an image generation model trained on datasets with limited diversity in human portraits may struggle to accurately represent a broad range of ethnicities. |

# III. IMPACTS OF BIAS IN AI

The rapid advancement of artificial intelligence (AI) has brought numerous benefits, but it also comes with potential risks and challenges. One of the key concerns is the negative impacts of bias in AI on individuals and society. Bias in AI can perpetuate and even amplify existing inequalities, leading to discrimination against marginalized groups and limiting their access to essential services. In addition to perpetuating gender stereotypes and discrimination, it can also lead to new forms of discrimination based on skin color, ethnicity, or physical appearance. To ensure that AI systems are fair, equitable, and serve the needs of all users, it is crucial to identify and mitigate bias in AI. Moreover, the use of biased AI has numerous ethical implications, including the potential for discrimination, responsibility of developers and policymakers, undermining public trust in technology, and limiting human agency and autonomy. Addressing these ethical implications will require a concerted effort from all stakeholders involved, and it is important to develop ethical guidelines and regulatory frameworks that promote fairness, transparency, and accountability in the development and use of AI systems.

## NEGATIVE IMPACTS OF BIAS IN AI ON INDIVIDUALS AND SOCIETY, INCLUDING DISCRIMINATION AND PERPETUATION OF EXISTING INEQUALITIES

The negative impacts of bias in AI can be significant, affecting individuals and society. Discrimination is a key concern when it comes to biased AI systems, as they can perpetuate and even amplify existing inequalities (Sweeney, 2013). For example, biased algorithms used in the criminal justice system can lead to unfair treatment of certain groups, particularly people of color, who are more likely to be wrongly convicted or receive harsher sentences (Angwin et al., 2016).

Bias in AI can also have a negative impact on individual's access to essential services, such as healthcare and finance. Biased algorithms can lead to underrepresentation of certain groups, such as people of color or those from lower socioeconomic backgrounds, in credit scoring systems, making it harder for them to access loans or mortgages (Dwork et al., 2012).

Furthermore, bias in AI can also perpetuate gender stereotypes and discrimination. For instance, facial recognition algorithms trained on data primarily consisting of men can struggle to recognize female faces accurately, perpetuating gender bias in security systems (Buolamwini & Gebru, 2018). When generative AI (GenAI) models are prompted to create images of CEOs, they tend to reinforce stereotypes by depicting CEOs predominantly as men (Nicoletti & Bass, 2023).

In addition to perpetuating existing inequalities, bias in AI can also lead to new forms of discrimination, such as those based on skin color, ethnicity, or even physical appearance. The same GenAI models that exhibit gender bias, perhaps unsurprisingly, also portray criminals or terrorists as people of color.

The public deployment of these systems can lead to serious consequences, such as denial of services, job opportunities, or even wrongful arrests or convictions. The risk is twofold: on an individual level, it affects people's perception of themselves and others, potentially influencing their opportunities and interactions.

On a societal level, the widespread use of such biased AI systems can entrench discriminatory narratives and hinder efforts toward equality and inclusivity. As AI becomes more integrated into our daily lives, the potential for such technology to shape cultural norms and social structures becomes more significant, making it imperative to address these biases in the developmental stages of AI systems to mitigate their harmful impacts (Ferrara, 2023; Ferrara, 2023b).

DISCUSSION OF THE ETHICAL IMPLICATIONS OF BIASED AI

The use of biased AI has numerous ethical implications that must be considered. One of the main concerns is the potential for discrimination against individuals or groups based on factors such as race, gender, age, or disability (Noble, 2018). When AI systems are biased, they can perpetuate existing inequalities and reinforce discrimination against marginalized groups. This is especially concerning in sensitive areas such as healthcare, where biased AI systems can lead to unequal access to treatment or harm patients (Obermeyer et al., 2019).

Another ethical concern is the responsibility of developers, companies, and governments in ensuring that AI systems are designed and used in a fair and transparent manner. If an AI system is biased and produces discriminatory outcomes, the responsibility lies not only with the system itself but also with those who created and deployed it (Mittelstadt et al., 2016). As such, it is crucial to establish ethical guidelines and regulatory frameworks that hold those responsible for the development and use of AI systems accountable for any discriminatory outcomes.

Moreover, the use of biased AI systems may undermine public trust in technology, leading to decreased adoption and even rejection of new technologies. This can have serious economic and social implications, as the potential benefits of AI may not be realized if people do not trust the technology or if it is seen as a tool for discrimination.

Finally, it is important to consider the impact of biased AI on human agency and autonomy. When AI systems are biased, they can limit individual freedoms and reinforce societal power dynamics. For example, an AI system used in a hiring process may disproportionately exclude candidates from marginalized groups, limiting their ability to access employment opportunities and contribute to society.

Addressing the ethical implications of biased AI will require a concerted effort from all stakeholders involved, including developers, policymakers, and society at large. It will be necessary to develop ethical guidelines and regulatory frameworks that promote fairness, transparency, and accountability in the development and use of AI systems (Ananny & Crawford, 2018). Additionally, it will be important to engage in critical discussions about the impact of AI on society and to empower individuals to participate in shaping the future of AI in a responsible and ethical manner.

## IV. MITIGATION STRATEGIES FOR BIAS IN AI

Researchers and practitioners have proposed various approaches to mitigate bias in AI. These approaches include pre-processing data, model selection, and post-processing decisions. However, each approach has its limitations and challenges, such as the lack of diverse and representative training data, the difficulty of identifying and measuring different types of bias, and the potential trade-offs between fairness and accuracy. Additionally, there are ethical considerations around how to prioritize different types of bias and which groups to prioritize in the mitigation of bias.

Despite these challenges, mitigating bias in AI is essential for creating fair and equitable systems that benefit all individuals and society. Ongoing research and development of mitigation approaches are necessary to overcome these challenges and ensure that AI systems are used for the benefit of all.

OVERVIEW OF CURRENT APPROACHES TO MITIGATE BIAS IN AI, INCLUDING PRE-PROCESSING DATA, MODEL SELECTION, AND POST-PROCESSING DECISIONS

Mitigating bias in AI is a complex and multifaceted challenge. However, several approaches have been proposed to address this issue. One common approach is to pre-process the data used to train AI models to ensure that it is representative of the entire population, including historically marginalized groups. This can involve techniques such as oversampling, undersampling, or synthetic data generation (Koh & Liang, 2017). For example, a study by Buolamwini and Gebru (2018) demonstrated that oversampling darker-skinned individuals improved the accuracy of facial recognition algorithms for this group. Pre-processing data involves identifying and addressing biases in the data before the model is trained. This can be done through techniques such as data augmentation, which involves creating synthetic data points to increase the representation of underrepresented groups, or through adversarial debiasing, which involves training the model to be resilient to specific types of bias (Zhang et al., 2018). Documenting such dataset biases and augmentation procedures is of paramount importance (Gebru et al., 2021).

Another approach to mitigate bias in AI is to carefully select the models used to analyze the data. Researchers have proposed using model selection methods that prioritize fairness, such as those based on group fairness (Yan et al., 2020) or individual fairness (Zafar et al., 2017). For example, a study by Kamiran and Calders (2012) proposed a method for selecting classifiers that achieve demographic parity, ensuring that the positive and negative outcomes are distributed equally across different demographic groups. Another approach is to use model selection techniques that prioritize fairness and mitigate bias. This can be done through techniques such as regularization, which penalizes models for making discriminatory predictions, or through ensemble methods, which combine multiple models to reduce bias (Dwork et al., 2018).

Post-processing decisions is another approach to mitigate bias in AI. This involves adjusting the output of AI models to remove bias and ensure fairness. For example, researchers have proposed post-processing methods that adjust the decisions made by a model to achieve equalized odds, which ensures that false positives and false negatives are equally distributed across different demographic groups (Hardt et al., 2016).

While these approaches hold promise for mitigating bias in AI, they also have limitations and challenges. For example, pre-processing data can be time-consuming and may not always be effective, especially if the data used to train models is already biased. Additionally, model selection methods may be limited by the lack of consensus on what constitutes fairness, and post-processing methods can be complex and require large amounts of additional data (Barocas & Selbst, 2016). Therefore, it is crucial to continue exploring and developing new approaches to mitigate bias in AI.

In the realm of generative AI, addressing bias is even more challenging as it requires a holistic strategy (Ferrara, 2023).This begins with the pre-processing of data to ensure diversity and representativeness. This involves the deliberate collection and inclusion of varied data sources that reflect the breadth of human experience, thus preventing the overrepresentation of any single demographic in training datasets. Model selection must then prioritize algorithms that are transparent and capable of detecting when they are generating biased outputs. Techniques such as adversarial training, where models are continually tested against scenarios designed to reveal bias, can be beneficial. Post-processing involves critically assessing the AI-generated content and, if necessary, adjusting the outputs to correct for biases. This might include using additional filters or transfer learning techniques to refine the models further. Regular audits, continuous monitoring, and incorporating feedback loops are essential to ensure that generative AI systems remain fair and equitable over time. These efforts must be underpinned by a commitment to ethical AI principles, actively engaging diverse teams in AI development, and fostering interdisciplinary collaboration to address and mitigate AI bias effectively.

Furthermore, implementing these approaches requires careful consideration of ethical and societal implications. For example, adjusting the model's predictions to ensure fairness may result in trade-offs between different forms of bias, and may have unintended consequences on the distribution of outcomes for different groups (Kleinberg et al., 2018; Ferrara, 2023c).

| Approach | Description | Examples | Limitations and Challenges | Ethical Considerations |
|---|---|---|---|---|
| **Pre-processing Data** | Involves identifying and addressing biases in the data before training the model. Techniques such as oversampling, undersampling, or synthetic data generation are used to ensure the data is representative of the entire population, including historically marginalized groups. | 1. Oversampling darker-skinned individuals in a facial recognition dataset (Buolamwini and Gebru, 2018). 2. Data augmentation to increase representation of underrepresented groups. 3. Adversarial debiasing to train the model to be resilient to specific types of bias (Zhang et al., 2018). | 1. Time-consuming process. 2. May not always be effective, especially if the data used to train models is already biased. | 1. Potential for over- or underrepresentation of certain groups in the data, which can perpetuate existing biases or create new ones. 2. Privacy concerns related to data collection and usage, particularly for historically marginalized groups. |
| **Model Selection** | Focuses on using model selection methods that prioritize fairness. Researchers have proposed methods based on group fairness or individual fairness. Techniques include regularization, which penalizes models for making discriminatory predictions, and ensemble methods, which combine multiple models to reduce bias. | 1. Selecting classifiers that achieve demographic parity (Kamiran and Calders, 2012). 2. Using model selection methods based on group fairness (Yan et al., 2020) or individual fairness (Zafar et al., 2017). 3. Regularization to penalize discriminatory predictions. 4. Ensemble methods to combine multiple models and reduce bias (Dwork et al., 2018). | Limited by the possible lack of consensus on what constitutes fairness. | 1. Balancing fairness with other performance metrics, such as accuracy or efficiency. 2. Potential for models to reinforce existing stereotypes or biases if fairness criteria are not carefully considered. |
| **Post-processing Decisions** | Involves adjusting the output of AI models to remove bias and ensure fairness. Researchers have proposed methods that adjust the decisions made by a model to achieve equalized odds, ensuring that false positives and false negatives are equally distributed across different demographic groups. | Post-processing methods that achieve equalized odds (Hardt et al., 2016). | Can be complex and require large amounts of additional data (Barocas & Selbst, 2016). | 1. Trade-offs between different forms of bias when adjusting predictions for fairness. 2. Unintended consequences on the distribution of outcomes for different groups. |

DISCUSSION OF THE LIMITATIONS AND CHALLENGES OF THESE APPROACHES
While various approaches have been proposed to address bias in AI, they also have their limitations and challenges.

One of the main challenges is the lack of diverse and representative training data. As mentioned earlier, data bias can lead to biased outputs from AI systems. However, collecting diverse and representative data can be challenging, especially when dealing with sensitive or rare events. Additionally, there may be privacy concerns when collecting certain types of data, such as medical records or financial information. These challenges can limit the effectiveness of dataset augmentation as a mitigation approach.

Another challenge is the difficulty of identifying and measuring different types of bias in AI systems. Algorithmic bias can be difficult to detect and quantify, especially when the algorithms are complex or opaque. Additionally, the sources of bias may be difficult to isolate, as bias can arise from multiple sources, such as the data, the algorithm, and the user. This can limit the effectiveness of bias-aware algorithms and user feedback mechanisms as mitigation approaches.

Moreover, mitigation approaches may introduce trade-offs between fairness and accuracy. For example, one approach to reducing algorithmic bias is to modify the algorithm to ensure that it treats all groups equally. However, this may result in reduced accuracy for certain groups or in certain contexts. Achieving both fairness and accuracy can be challenging and requires careful consideration of the trade-offs involved.

Finally, there may be ethical considerations around how to prioritize different types of bias and which groups to prioritize in the mitigation of bias. For example, should more attention be paid to bias that affects historically marginalized groups, or should all types of bias be given equal weight? These ethical considerations can add complexity to the development and implementation of bias mitigation approaches.

Despite these challenges, addressing bias in AI is crucial for creating fair and equitable systems. Ongoing research and development of mitigation approaches is necessary to overcome these challenges and to ensure that AI systems are used for the benefit of all individuals and society.

# V. FAIRNESS IN AI

Fairness in AI is a critical topic that has received a lot of attention in both academic and industry circles. At its core, fairness in AI refers to the absence of bias or discrimination in AI systems, which can be challenging to achieve due to the different types of bias that can arise in these systems. There are several types of fairness proposed in the literature, including group fairness, individual fairness, and counterfactual fairness. While fairness and bias are closely related concepts, they differ in important ways, including that fairness is inherently a deliberate and intentional goal, while bias can be unintentional. Achieving fairness in AI requires careful consideration of the context and stakeholders involved. Real-world examples of fairness in AI demonstrate the potential benefits of incorporating fairness into AI systems.

DEFINITION OF FAIRNESS IN AI AND ITS DIFFERENT TYPES
Fairness in AI is a complex and multifaceted concept that has been the subject of much debate in both the academic and industry communities. At its core, fairness refers to the absence of bias or discrimination in AI systems (Barocas & Selbst, 2016). However, achieving fairness in AI can be challenging, as it requires careful consideration of the different types of bias that can arise in these systems and the ways in which they can be mitigated.

There are several different types of fairness that have been proposed in the literature, including group fairness, individual fairness, and counterfactual fairness (Zafar et al., 2017).

Group fairness refers to ensuring that different groups are treated equally or proportionally in AI systems. This can be further subdivided into different types, such as demographic parity, which ensures that the positive and negative outcomes are distributed equally across different demographic groups (Kamiran & Calders, 2012), a notion of unfairness, disparate mistreatment, defined in terms of misclassification rates (Zafar et al., 2017), or equal opportunity, which ensures that the true positive rate (sensitivity) and false positive rate (1-specificity) are equal across different demographic groups (Hardt et al., 2016).

Individual fairness, on the other hand, refers to ensuring that similar individuals are treated similarly by AI systems, regardless of their group membership. This can be achieved through methods such as similarity-based or distance-based measures, which aim to ensure that individuals who are similar in terms of their characteristics or attributes are treated similarly by the AI system (Dwork et al., 2012).

Counterfactual fairness is a more recent concept that aims to ensure that AI systems are fair even in hypothetical scenarios. Specifically, counterfactual fairness aims to ensure that an AI system would have made the same decision for an individual, regardless of their group membership, even if their attributes had been different (Kusner et al., 2017).

Other types of fairness include procedural fairness, which involves ensuring that the process used to make decisions is fair and transparent, and causal fairness, which involves ensuring that the system does not perpetuate historical biases and inequalities (Kleinberg et al., 2018).

It is important to note that these different types of fairness are not mutually exclusive and may overlap in practice. Additionally, different types of fairness may conflict with each other, and trade-offs may need to be made to achieve fairness in specific contexts (Barocas & Selbst, 2016). It is important to note that achieving fairness in AI is not a one-size-fits-all solution and requires careful consideration of the context and stakeholders involved. Achieving fairness in AI systems often requires a nuanced understanding of these different types of fairness and the ways in which they can be balanced and prioritized in different contexts.

## COMPARISON OF FAIRNESS AND BIAS IN AI

While fairness and bias are closely related concepts, they differ in important ways. Bias refers to the systematic and consistent deviation of an algorithm's output from the true value or from what would be expected in the absence of bias (Zliobaite, 2021). On the other hand, fairness in AI refers to the absence of discrimination or favoritism towards any individual or group based on their protected characteristics such as race, gender, age, or religion (Dwork et al., 2012).

One key difference between fairness and bias is that while bias can be unintentional, fairness is inherently a deliberate and intentional goal. Bias can arise due to various factors, such as biased data or algorithmic design, but fairness requires a conscious effort to ensure that the algorithm does not discriminate against any group or individual. In other words, bias can be viewed as a technical issue, while fairness is a social and ethical issue (Barocas & Selbst, 2016).

Another difference is that bias can be either positive or negative, whereas fairness is only concerned with negative bias or discrimination (Zliobaite, 2021). Positive bias occurs when an algorithm systematically favors a particular group or individual, while negative bias occurs when the algorithm systematically discriminates against a particular group or individual. In contrast, fairness is concerned with preventing negative bias or discrimination towards any group or individual.

Despite these differences, fairness and bias are often closely related, and addressing bias is an important step towards achieving fairness in AI. For example, addressing bias in training data or algorithms can help reduce the likelihood of unfair outcomes. However, it is important to recognize that bias is not the only factor that can lead to unfairness, and achieving fairness may require additional efforts beyond bias mitigation (Kleinberg et al., 2017).

Overall, understanding the differences between fairness and bias is important for developing effective strategies to mitigate bias and ensure fairness in AI systems. By acknowledging these differences and designing algorithms and systems that prioritize fairness, we can ensure that AI systems are used to benefit all individuals and groups, without perpetuating or exacerbating existing social and economic inequalities.

| Type of Fairness | Description | Examples |
| --- | --- | --- |
| **Group Fairness** | Ensures that different groups are treated equally or proportionally in AI systems. Can be further subdivided into demographic parity, disparate mistreatment, or equal opportunity. | 1. Demographic parity: Positive and negative outcomes distributed equally across demographic groups (Kamiran & Calders, 2012). 2. Disparate mistreatment: Defined in terms of misclassification rates (Zafar et al., 2017). 3. Equal opportunity: True positive rate (sensitivity) and false positive rate (1-specificity) are equal across different demographic groups (Hardt et al., 2016). |
| **Individual Fairness** | Ensures that similar individuals are treated similarly by AI systems, regardless of their group membership. Can be achieved through methods such as similarity-based or distance-based measures. | Using similarity-based or distance-based measures to ensure that individuals with similar characteristics or attributes are treated similarly by the AI system (Dwork et al., 2012). |
| **Counterfactual Fairness** | Aims to ensure that AI systems are fair even in hypothetical scenarios. Specifically, counterfactual fairness aims to ensure that an AI system would have made the same decision for an individual, regardless of their group membership, even if their attributes had been different. | Ensuring that an AI system would make the same decision for an individual, even if their attributes had been different (Kusner et al., 2017). |
| **Procedural Fairness** | Involves ensuring that the process used to make decisions is fair and transparent. | Implementing a transparent decision-making process in AI systems. |
| **Causal Fairness** | Involves ensuring that the system does not perpetuate historical biases and inequalities. | Developing AI systems that avoid perpetuating historical biases and inequalities (Kleinberg et al., 2018). |

## REAL-WORLD EXAMPLES OF FAIRNESS IN AI

There have been various real-world examples of fairness in AI that demonstrate the potential benefits of incorporating fairness into AI systems. One example is the COMPAS (Correctional Offender Management Profiling for Alternative Sanctions) system, which is used to predict the likelihood of recidivism in criminal defendants. Research has shown that the system was biased against African American defendants, as it was more likely to falsely predict that they would reoffend compared to white defendants (Angwin et al., 2016). To address this issue, the Northpointe COMPAS was modified to include a "race-neutral" version of the algorithm that achieved similar accuracy while reducing racial bias (Larson et al., 2016).

Another example is the use of AI in the recruitment process. Research has shown that AI recruitment systems can be biased against women, as they may be less likely to be selected for male-dominated roles (Dastin, 2018). To address this issue, some companies have implemented "gender decoder" tools that analyze job postings and suggest changes to reduce gender bias (Crawford, 2019).

A third example is the use of AI in healthcare. Research has shown that AI systems used to predict healthcare outcomes can be biased against certain groups, such as African Americans (Obermeyer et al., 2019). To address this issue, researchers have proposed using techniques such as subgroup analysis to identify and address bias in the data used to train AI models (Lamy et al., 2020).

These real-world examples demonstrate the potential benefits of incorporating fairness into AI systems. By addressing bias and ensuring fairness, AI systems can be more accurate, ethical, and equitable, and can help to promote social justice and equality.

# VI. Mitigation Strategies for Fairness in AI

As the use of artificial intelligence (AI) continues to grow, ensuring fairness in its decision-making is becoming increasingly important. The use of AI in critical domains such as healthcare, finance, and law have the potential to significantly impact people's lives, and therefore, it is crucial that these systems make fair and unbiased decisions. To address this challenge, various approaches have been developed, including group fairness and individual fairness. However, these approaches are not without limitations and challenges, such as trade-offs between different types of fairness and the difficulty of defining fairness itself. In this section, we will explore mitigation strategies for fairness in AI, including current approaches, challenges, and areas for future research. By developing a better understanding of these mitigation strategies, we can work towards creating AI systems that are fair, unbiased, and equitable for all.

## OVERVIEW OF CURRENT APPROACHES TO ENSURE FAIRNESS IN AI, INCLUDING GROUP FAIRNESS AND INDIVIDUAL FAIRNESS

Ensuring fairness in AI is a complex and evolving field, with various approaches being developed to address different aspects of fairness. Two key approaches that have emerged are group fairness and individual fairness.

Group fairness is concerned with ensuring that AI systems are fair to different groups of people, such as people of different genders, races, or ethnicities. Group fairness aims to prevent the AI system from systematically discriminating against any group. This can be achieved through various techniques such as re-sampling, pre-processing, or post-processing the data used to train the AI model. For example, if an AI model is trained on data that is biased towards a particular group, re-sampling techniques can be used to create a balanced dataset, where each group is represented equally. Other techniques, such as pre-processing or post-processing, can be used to adjust the output of the AI model to ensure that it does not unfairly disadvantage any group. Corbett-Davies and collaborators introduced risk-minimization approaches aimed to minimize disparities (Corbett-Davies et al., 2017; Corbett-Davies & Goel, 2018).

Individual fairness, on the other hand, is concerned with ensuring that AI systems are fair to individuals, regardless of their group membership. Individual fairness aims to prevent the AI system from making decisions that are systematically biased against certain individuals. Individual fairness can be achieved through techniques such as counterfactual fairness or causal fairness. For example, counterfactual fairness aims to ensure that the AI model would have made the same decision for an individual, regardless of their race or gender.

While group fairness and individual fairness are important approaches to ensuring fairness in AI, they are not the only ones. Other approaches include transparency, accountability, and explainability. Transparency involves making the AI system's decision-making process visible to users, while accountability involves holding the system's developers responsible for any harm caused by the system. Explainability involves making the AI system's decisions understandable to users (Donovan et al., 2018; Ananny & Crawford,. 2018).

Overall, ensuring fairness in AI is a complex and ongoing challenge that requires a multi-disciplinary approach, involving experts from fields such as computer science, law, ethics, and social science. By developing and implementing a range of approaches to ensure fairness, we can work towards creating AI systems that are unbiased, transparent, and accountable.

## DISCUSSION OF THE LIMITATIONS AND CHALLENGES OF THESE APPROACHES

While these approaches have shown promising results in promoting fairness in AI, they are not without limitations and challenges. One major limitation is the potential for trade-offs between different types of fairness. For example, group fairness approaches may result in unequal treatment of individuals within a group, while individual fairness approaches may not address systemic biases that affect entire groups (Barocas & Selbst, 2016). Additionally, it may be difficult to determine which types of fairness are most appropriate for a given context, and how to balance them appropriately (Kleinberg et al., 2018).

Another challenge is the difficulty of defining fairness itself. Different people and groups may have different definitions of fairness, and these definitions may change over time (Dwork et al., 2018). This can make it challenging to develop AI systems that are considered fair by all stakeholders.

Furthermore, many of the current approaches to ensuring fairness in AI rely on statistical methods and assumptions that may not accurately capture the complexity of human behavior and decision-making. For example, group fairness metrics may not consider intersectionality, or the ways in which different dimensions of identity (such as race, gender, and socioeconomic status) interact and affect outcomes (Crenshaw. 1989).

Finally, there are concerns about the potential for unintended consequences and harmful outcomes resulting from attempts to ensure fairness in AI. For example, some researchers have found that attempts to mitigate bias in predictive policing algorithms may increase racial disparities in arrests (Ferguson, 2012).

Despite these challenges, the development of fair and equitable AI is an important and ongoing area of research. Future work will need to address these challenges and continue to develop new approaches that are sensitive to the nuances of fairness and equity in different contexts.

| Approach | Description | Examples | Limitations and Challenges |
|---|---|---|---|
| **Group Fairness** | Ensures that AI systems are fair to different groups of people, such as people of different genders, races, or ethnicities. Aims to prevent the AI system from systematically discriminating against any group. Can be achieved through techniques such as re-sampling, pre-processing, or post-processing the data. | 1. Re-sampling techniques to create a balanced dataset. 2. Pre-processing or post-processing to adjust AI model output. | 1. May result in unequal treatment of individuals within a group. 2. May not address systemic biases that affect individual characteristics. 3. Group fairness metrics may not consider intersectionality. |
| **Individual Fairness** | Ensures that AI systems are fair to individuals, regardless of their group membership. Aims to prevent the AI system from making decisions that are systematically biased against certain individuals. Can be achieved through techniques such as counterfactual fairness or causal fairness. | 1. Counterfactual fairness ensuring the same decision regardless of race or gender. | 1. May not address systemic biases that affect entire groups. 2. Difficulty determining which types of fairness are appropriate for a given context and how to balance them. |
| **Transparency** | Involves making the AI system's decision-making process visible to users. | Making AI system's decisions and processes understandable to users. | Different definitions of fairness among people and groups, and changing definitions over time. |
| **Accountability** | Involves holding the system's developers responsible for any harm caused by the system. | Developers held responsible for unfair decisions made by AI systems. | Determining responsibility and addressing potential harm. |
| **Explainability** | Involves making the AI system's decisions understandable to users. | Providing clear explanations of AI system's decisions. | Addressing the complexity of human behavior and decision-making. |
| **Intersectionality** *(not explicitly mentioned as an approach, but it is an aspect to consider)* | Considers the ways in which different dimensions of identity (such as race, gender, and socioeconomic status) interact and affect outcomes. | Developing AI systems that consider the interaction of different dimensions of identity. | Addressing the complexity of intersectionality and ensuring fairness across multiple dimensions of identity. |

# VII. CONCLUSIONS

In conclusion, this paper has illuminated the various sources of biases in AI and ML systems and their profound societal impact, with an extended discussion on the emergent concerns surrounding generative AI bias. It is clear that these powerful computational tools, if not diligently designed and audited, have the potential to perpetuate and even amplify existing biases, particularly those related to race, gender, and other societal constructs. We have considered numerous examples of biased AI systems, with a particular focus on the intricacies of generative AI, which illustrates the critical need for comprehensive strategies to identify and mitigate biases across the entire spectrum of the AI development pipeline.

To combat bias, this paper has highlighted strategies such as robust data augmentation, the application of counterfactual fairness, and the imperative for diverse, representative datasets alongside unbiased data collection methods. We also considered the ethical implications of AI in preserving privacy and the necessity for transparency, oversight, and continuous evaluation of AI systems.

As we look to the future, research in fairness and bias in AI and ML should prioritize the diversification of training data and address the nuanced challenges of bias in generative models, especially those used for synthetic data creation and content generation. It is imperative to develop comprehensive frameworks and guidelines for responsible AI and ML, which includes transparent documentation of training data, model choices, and generative processes. Diversifying the teams involved in AI development and evaluation is equally crucial, as it brings a multiplicity of perspectives that can better identify and correct for biases (Stathoulopoulos & Mateos-Garcia, 2019).

Lastly, the establishment of robust ethical and legal frameworks governing AI and ML systems is paramount, ensuring that privacy, transparency, and accountability are not afterthoughts but foundational elements of the AI development lifecycle (Wachter et al., 2018). Research must also explore the implications of generative AI, ensuring that as we advance in creating ever more sophisticated synthetic realities, we remain vigilant and proactive in safeguarding against the subtle encroachment of biases that could shape society in unintended and potentially harmful ways.

## ACKNOWLEDGEMENTS

The author is indebted to all current and past members of his lab at USC, core researchers and visiting students at ISI, collaborators and coauthors of work related to AI and fairness.